\magnification=\magstep1
\dimen100=\hsize
\parskip= 6pt

\font\ninerm=cmr10 at 10truept
\font\eightrm=cmr8
\font\sc=cmcsc10

\font\tenmsy=msbm10
\font\sevenmsy=msbm7
\font\fivemsy=msbm5
\newfam\msyfam
\def\msy{\fam\msyfam\tenmsy}
\textfont\msyfam=\tenmsy
\scriptfont\msyfam=\sevenmsy
\scriptscriptfont\msyfam=\fivemsy

\def\bbc{{\msy C}}

\def\bbh{{\msy H}}
\def\bbi{{\msy I}}

\def\bbp{{\msy P}}

\def\bbz{{\msy Z}}
\def\za{\vrule height6pt width4pt depth1pt}

\def \otimescomma {\ {\scriptstyle {{}_\otimes\atop {{}^,}}}}

\def\Ra*{(R_{a})_{*}}
\def\ada-1{ad_{a^{-1}}~}
\def\p-1u{\pi^{-1}(U)}

\font\svtnrm=cmr17

\def \mrat {{\cal M}_{\rm rat}}
\def \mell {{\cal M}_{\rm ell}}
\def \mtrig {{\cal M}_{\rm trig}}

\centerline{\svtnrm Multi-Hamiltonian structures for r-matrix systems} 
\bigskip
\centerline{\sc  J. Harnad and J.C. Hurtubise}
\bigskip
\centerline{CRM preprint 2850}

\footnote{}{\ninerm The authors of this article would like to thank 
NSERC and FCAR for their support}
\bigskip
\centerline{\vbox{\hsize = 5.85truein
\baselineskip = 12.5truept
\eightrm
{\sc Abstract}: For the rational, elliptic and trigonometric r-matrices, we
 exhibit the links between three ``levels" of Poisson spaces: (a) Some 
finite-dimensional
spaces of matrix-valued 
holomorphic functions on the complex line; (b) Spaces of spectral curves 
and sheaves supported on them;
(c) Symmetric products of a surface. We have, at each level, a linear 
space of 
compatible Poisson structures, and the maps relating the levels are 
Poisson. This leads in a natural way to 
Nijenhuis coordinates for these spaces.  At level (b), there are Hamiltonian systems on 
these spaces which are integrable for each Poisson structure in the family, 
and which are such that the Lagrangian leaves are the intersections of the 
symplective leaves over the Poisson structures in the family. Specific 
examples include many of the well-known integrable systems.}}

\noindent{\bf 1. Introduction} 

In this note, we show that a linear (``compatible'') family 
of holomorphic Poisson 
brackets defined on certain subspaces of the loop algebra of $Gl(r,\bbc)$ 
(``level (a)'')
endows them with a  multi-Hamiltonian
structure (see, e.g. Magri [Mg1] ), and that it
possesses a common set of Nijenhuis or Darboux-Nijenhuis coordinates (in  a  sense to 
be defined below). This family of brackets
contains both the linear Lie Poisson structures on the algebra, as well
as the quadratic structures due to Sklyanin [Sk1,2]. The brackets are all
defined in terms of $r$-matrices, and we shall consider three cases, 
the rational, trigonometric and elliptic $r$-matrices.

This family of subspaces, with their accompanying   families of brackets 
and
Hamiltonians,   is an important 
one, since it 
includes many, if not  most, of the well-known cases of integrable 
systems:
to cite  some familiar examples, the Gaudin model, spin chains, the Toda lattice, the 
various tops (Euler, Lagrange, Manakov, Kovalevski, Steklov),
 the Landau Lifschitz equation, as well as the finite gap cases of the KdV, 
NLS, CNLS or Boussinesq hierarchies.
 References include  the book [FT], the survey [RS2] and the references
therein,  or the articles [Mo, AvM, RS1, AHP, 
HHu].

In the  three cases of rational, elliptic and trigonometric $r$-matrices, 
the phase spaces
 have	geometric 
interpretations as spaces of meromorphic endomorphisms of a vector bundle 
over
a compact Riemann surface. This bundle has the property that it
 is holomorphically rigid under deformation, at least if one fixes the top 
exterior
power. Bundles with this property exist on the Riemann sphere
(rational case), elliptic curves (elliptic case) and their
nodal degenerations (trigonometric case). This interpretation will be crucial in the 
geometric study of the systems.

In all three cases, once one reduces by	the 
action of the group of automorphisms of the bundle,
one can describe the reduced phase space as a space
of pairs (spectral curve, line bundle on the spectral curve) (``level 
(b)''). 
These spaces, in turn,
 have natural families of Poisson brackets on them, the Mukai brackets, which turn 
out to be identical to the (reduced) brackets of our family. 

Furthermore, each of these spaces admits Poisson maps to
a symmetric product of a surface equipped with a family of non-degenerate
Poisson structures (``level(c)''). (More properly, one should consider a 
Hilbert scheme  instead of a symmetric product). The point of this note is that all 
three levels possess 
linear families of Poisson structures, and these families can be 
identified in such a way that
the maps relating the three levels are all Poisson, for any Poisson 
structure in the family.
In particular, the surface of level (c) provides the Nijenhuis coordinates 
referred to above.

In short we will prove the following 

{\sc Theorem} 1. \tensl For the rational, trigonometric and elliptic 
$r$-matrices, there are maps from 
\item {(a)}  A space $\cal M$ of, in the rational case,  elements of the loop algebra on 
$Gl(r,\bbc)$, and in the trigonometric and elliptic case, of pairs (element of the loop algebra on 
$Gl(r,\bbc)$, point on a fixed nodal rational or elliptic curve)
\item {(b)} A space $\cal N$ of pairs (spectral curves $S$, torsion free 
sheaves $L$ on $S$), to 
\item {(c)} The Hilbert scheme $Hilb^g(T)$ of a surface $T$, where $g$ = 
genus ($S$).

Each level posseses a linear family of Poisson structures, and the maps are 
Poisson for each of these structures. The maps from level (a) to (b) are 
quotients by a group of automorphisms. The maps from level (b) to (c) are 
generically immersions on symplectic leaves. At level (c), the Poisson 
structures are all non-degenerate on open sets, and there are natural
 coordinates  which provide Nijenhuis coordinates for the system. \tenrm

Precise definitions are given below. One has a natural set of  
Hamiltonians  defined at level (a), given in terms of the characteristic 
polynomial of the matrices, and equivalently in terms of the 
coefficients of the equation of the spectral curve $S$.
We will show 

{\sc Theorem} 2. \tensl The map $(S,L) \mapsto S$  defines an integrable system, 
that is a Lagrangian fibration, on the spaces $\cal N$, for each Poisson 
structure in the family at level (b). The Lagrangian leaves are cut out by 
fixing the values of the Casimirs for each structure in the family. \tenrm

As one varies the Poisson structure, then,  some of the Casimirs for 
one structure turn into ``effective" 
Hamiltonians for another of the structures (i.e., giving actual flows),
 in such a way that the union 
over the Poisson structures of the Casimirs generates the ring of 
Hamiltonians which Poisson commutes for any of the structures. This can be 
thought of as an example of a generalization of the Gel'fand Zakharevich theorem [GZ].
In section 2, we will give the three levels of phase space that
we consider, for each of the three cases (rational, elliptic, 
trigonometric)
that we are considering,
and state the theorems that relate them. In section 3, we
exhibit the multi-Hamiltonian structure, and the Nijenhuis coordinates. 
Section 4 is devoted to making some of the formulae explicit in the 
rational case, as well as some examples.

We would like to thank Eyal Markman and Franco Magri for useful 
discussions.

\noindent {\bf 2. Phase spaces.}

\noindent{\sc a) The Lie-Poisson-Sklyanin phase spaces} 

{\it i) Rational case}

For this case, fixing an integer $n$, the phase space is simply 
$$\mrat = \{ gl(r,\bbc)-{\rm valued\ polynomials\ 
of\ degree\ } \le n\}.\eqno(2.1)$$
Let $\Sigma$ be $\bbp_1$, the Riemann sphere. Let ${\cal O}(m)$ denote the 
standard degree $m$ line bundle on $\Sigma$, and set $F(m) = F\otimes
{\cal O}(m)$ for any vector bundle $F$. Let $K_\Sigma= {\cal O}(-2)$ be the 
canonical
(=cotangent) bundle of $\Sigma$. If $E$ is the trivial rank $r$
vector bundle on $\Sigma$, $\mrat$ can be reinterpreted as:
$$\mrat = H^0(\Sigma, End(E)(n)) = H^0(\Sigma, End(E)\otimes 
K_\Sigma(n+2)),\eqno(2.2)$$
that is the global holomorphic sections of the endomorphisms of $E$ with a pole of 
order $n$
at infinity, or alternately,
of  the 1-form-valued endomorphisms of $E$ with a pole of 
degree $n+2$ at infinity.

The dual space to $\mrat$, using the trace residue pairing
$ <a,b> = tr(res_\infty(ab))$,   can be identified with 
the space of matricial Laurent polynomials with entries of degree 
$-n-1,..., -1$.
For any pair $f,g$ of functions on $\mrat$, therefore, the differentials $df, dg$ at 
a point
may be identified as such Laurent polynomials.
In a more invariant fashion, these Laurent polynomials are representative 
cocycles for cohomology classes, and the dual space of $\mrat$ is 
$$\mrat^* = H^1(\Sigma, End(E)(-n-2)),\eqno(2.3)$$
This is just Serre duality.
 Let $P_+$ be the projection defined on the space 
of all Laurent polynomials which is the identity on the terms of degree greater or 
equal to zero, and sets to zero the terms of strictly negative degree.  Let 
$P_-$ be the complementary projection:
$P_+ + P_- = \bbi$. We set 
$$R= P_+-P_-.\eqno(2.4)$$
On the space $\mrat$, one has an $(n+3)$ dimensional family of Poisson 
structures defined 
as follows: let $a = a(\lambda) $ be a polynomial of degree at most $n+1$, 
and $b$
 be a constant.
 Our Poisson structures at $\phi\in \mrat$ are given by:
$$\{f,g\}(\phi) = <\phi, [R(a\ df) , dg] +[df, R(a\ dg)]> -{b\over 2}(<R( Df), 
Dg> +  <D'f,R(D'g)>),\eqno(2.5)$$
where $D$ denotes the left derivative and $D'$ the right derivative. These  
are defined 
on tangent vectors $\dot \phi$ at $\phi$ by 
$$< Df, \dot \phi \phi^{-1}> = <df,\dot \phi> = <D'f,\phi^{-1}\dot 
\phi>,\eqno(2.6)$$
so that $Df = \phi\ df, D'f = df\ \phi.$

One can write the Poisson bracket in a different fashion. Noting that 
the Poisson brackets
are determined by their values on the matrix entries $\phi_{i,j}(\lambda)$ as 
$i,j,\lambda$ vary,
and   that the projection $P_+(f) (\lambda)$ of a function $f$ can be 
defined by the 
contour integral
$$P_+(f) (\lambda) = {1\over 2\pi i } \oint {1\over \mu-\lambda} f(\mu)d\mu,\eqno(2.7)$$
we can obtain, after some computations:
$$
\{\phi(\lambda) \otimescomma  \phi(\mu)\}_{a,b} :=
[r(\lambda-\mu), \ \phi(\lambda) \otimes (a(\mu)\bbi -{b\over 2} \phi(\mu))
   + (a(\lambda)\bbi -{b\over 2} \phi(\lambda))\otimes \phi(\mu)],
 \eqno(2.8)$$
Here we use  the tensor-bracket notation of [FT], considering both sides 
as elements of $End (\bbc^r\otimes\bbc^r)$; $r(\lambda-\mu)$ is the 
explicit expression of the $r$-matrix.  When $b=0$, the Poisson structure is equivalent
 to the standard linear $r$-matrix bracket on the family of matrices of 
the form
$a(\lambda)^{-1}\phi, \phi\in \mrat$; when $a=0, b=-1$, the bracket is the 
standard quadratic
(Sklyanin) bracket.

{\it ii) Elliptic case: (cf. [FT]) }

Let $\Sigma$ be an elliptic curve, defined as
$$\Sigma = \bbc/(\omega_1\bbz+\omega_2\bbz), \eqno(2.9)$$
and let $\pi:\bbc\rightarrow \Sigma$ be the natural projection.

Let $q=	{\rm exp} (2\pi i/r)$, and set
$$I_1 ={\rm diag} (1,q, q^2,...,q^{r-1}),\quad\quad
I_2= \pmatrix {0&1&0&\dots&0\cr 0&0&1&\dots&0\cr.&.&.& &.\cr.&.&.& &.\cr
0&0&0&\dots&1\cr1&0&0&\dots&0}.\eqno(2.10)$$
Note that $I_1I_2I_1^{-1}I_2 ^{-1}= q^{-1}\bbi . $

Let $D$ be a divisor $\nu_1+\nu_2+...+\nu_n$ on $\Sigma$, so that the	
$\nu_i$ are points (possibly repeated)  on $\Sigma$.

Our phase space $\mell$ will be the product of the curve with the space of 
meromorphic functions on $\bbc$
with values in $gl(r,\bbc)$, with poles only at the translates of the 
$\nu_i$ and
satisfying the quasiperiodicity relations: 
$$\eqalign{\mell = \Sigma \times \{ \phi  &{\rm \ meromorphic}\ 
gl(r,\bbc)-{\rm valued\ functions\ on\ }\bbc\ {\rm\ such\ that}\cr
           & \phi^{-1}(\infty)\subset 
\cup_i\{\nu_i+\bbz\omega_1+\bbz \omega_2\} 
  {\rm\ and\ } \phi(\lambda+\omega_i) = I_i\phi (\lambda)I_i^{-1}, i=1,2.\} 
}\eqno(2.11)$$

The second factor can again be thought of as a space of meromorphic 
sections of the endomorphism bundle 
 of a vector bundle, this time  of degree 1. The first factor is to be 
thought of
as classifing the bundle. Indeed, referring to  Atiyah
[At],	stable vector bundles $E$ of degree 1, 
rank $r$, are classified by their top exterior power
$\Lambda^r(E)\in Pic^1(\Sigma) = \Sigma$, and are all related to each 
other by tensoring by a line bundle: $E' = E\times L$, so that in particular
$End(E') = End(E)$. They all have a one-dimensional 
space of sections.
 By the theorem of Narasimhan and Seshadri[NS], 
stable bundles correspond to	
irreducible representations of 
a $\bbz$-central extension of the
fundamental group, in our case given by the matrices $(\tilde I_1, \tilde 
I_2)= (I_1, cI_2)$, for a 
constant 
$c$. Sections of the 
corresponding vector bundles $E_c$ correspond to vector valued functions
on $\bbc$ satisfying the quasi-periodicity relations $ \psi(\lambda+\omega_i) = 
\tilde I_i\psi (\lambda)$. 
Sections of $End(E_c)$ in turn, are given by holomorphic 
matrix valued functions on $\bbc$, satisfying the 
quasi-periodicity 
relations of (2.11) on $\bbc$. Setting $E= E_1$, the space $\mell$ then gets 
interpreted as 
  sections of $End(E)$, meromorphic, with poles at $D$:
$$\mell = \Sigma \times H^0(\Sigma,End(E)(D)) =	\Sigma \times H^0(\Sigma, 
End(E)\otimes K_\Sigma(D)).\eqno(2.12)$$
Corresponding to the fact that the vector bundle $E$ is rigid once one 
fixes the top exterior
power, one has $H^1(\Sigma, sl(E) )= 0$, and dually, $H^0(\Sigma, sl(E) )= 
0$. 
This, in concrete terms, means that any section $s$ of 
$sl(E)$ on a punctured neighbourhood of the origin in $\Sigma$ decomposes 
uniquely into 
$s_+ + s_-$, where $s_+$ is a section defined on an unpunctured 
neighbourhood $U_+$ of the origin, and
$s_-$ is defined on $U_- = \Sigma - \{{\rm origin}\}$. Set $P_+(s)= s_+, 
P_-(s)= s_-$. Using the 
representation of sections of $E$ given above, another way of giving this 
decomposition is to say that 
any meromorphic $sl(r,\bbc)$-valued function $f$ on a  punctured 
neighbourhood of the origin 
in $\Sigma$ decomposes uniquely into 
$f_+ + f_-$, where $f_+$ is defined on an unpunctured neighbourhood of the 
origin in $\bbc$, and
$f_-$ is defined on $\bbc-\{{\rm translates\ of\ the\ origin}\}$, and 
satisfies the 
quasi-periodicity relations  of (2.11). Again, we denote the corresponding 
projections by $P_+, P_-$.

More generally, a section $\rho$ of $End(E)$ on a punctured neighbourhood of the 
origin in $\Sigma$ can be decomposed   into its trace component
$P_0(\rho)$  and its $sl(r,\bbc)$ component $\rho-P_0(\rho)$; in turn, the 
latter decomposes
 into its components $P_+(\rho) = P_+(\rho-P_0(\rho))$,
$P_-(\rho) = P_-(\rho-P_0(\rho))$, so that $\rho= P_+(\rho) + P_-(\rho) + 
P_0(\rho)$. As before, set $R = P_+ - P_-$.

The cotangent space of $\mell$ at $(E_c, \phi)$ is identified with 
$\bbc \times H^1(\Sigma,End(E)(-D))$: the differential $df$ of a 
function $f$ on $\mell$ splits into $df_E\in \bbc$, $df_\phi\in 
H^1(\Sigma,End(E)(-D))$.  
The line bundle ${\cal O}(D)$ on $\Sigma$, whose sections correspond to 
meromorphic functions
 with poles only at $D$,
 has an $n$-dimensional space $V$ of global sections. If $a\in V$, 
 representing $df_\phi$ by  a cocycle one has 
 that $a\ df_\phi$ is a cocycle with values in $End(E)$, and so can be 
 split into its $+, -, 0$ components; similarly, $Df_\phi = \phi\ df_\phi, 
 D'f_\phi = df_\phi\ \phi$ also lie in $End(E)$, and can also be split.

 One can define an
$n+1$-dimensional	family 
of Poisson structures on $\mell$ by 
$$\eqalign { \{f,g\}(\phi) &= <\phi, [R(a\ df_\phi) , dg_\phi] +[df_\phi, 
R(a\ dg_\phi)]> 
\cr & \quad -{b \over 2}(<R(Df_\phi), Dg_\phi> + <D'f_\phi, R(D'g_\phi)>) 
\cr
&\quad + <P_0(a df_\phi +bDf_\phi), dg_E>- <df_E,  P_0((a  
dg_\phi+bDg_\phi)>,}\eqno(2.13)$$
 where $a$ lies in $V$, and $b$ is a constant.  The case $b=0$ again 
corresponds to the linear $r$-matrix bracket; the case $a= 0$
gives the quadratic bracket. One can again write out a formula as in (2.8) 
for the bracket, this time involving elliptic functions.

{\it Reduction to $sl(r)$}

The more familiar form of either the linear or quadratic   elliptic 
$r$-matrix bracket lives
on $sl(r,\bbc)$-valued functions, in the linear case, or on 
$Sl(r,\bbc)$-valued functions,
in the quadratic case. In our context, these arise by first fixing the values of some 
Casimirs, then reducing. 
The projection to the first factor in $\mell$ gives a Hamiltonian 
function, which acts 
on the  pairs $(c, \phi)$ by 
$$(\dot c,\dot  \phi) = (0, a+b\phi).\eqno(2.14)$$
Reducing then amounts to fixing the first factor in $\mell$ (i.e., fixing 
the bundle),
and then quotienting by  the flow (2.14).

When $b=0$, the flow acts by adding to 
$\phi$ a multiple of the 
identity. The polar parts of the trace components are Casimirs; we set 
these to zero. The easy
 normalisation for quotienting by the flow is given by fixing the constant
 term of the trace, and so we set this to zero also,  so that the whole trace
 term vanishes. We are now in 
$sl(r,\bbc)$, with a reduced phase space
$$\eqalign{(\mell)_{\rm red} =  \{ \phi  &{\rm \ meromorphic}\ 
sl(r,\bbc)-{\rm valued\ functions\ on\ }\bbc\ {\rm\ such\ that}\cr
           & \phi^{-1}(\infty)\subset 
\cup_i\{\nu_i+\bbz\omega_1+\bbz \omega_2\} 
  {\rm\ and\ } \phi(z+\omega_i) = I_i\phi (z)I_i^{-1}, i=1,2\}. 
}\eqno(2.15)$$

Similarily, when $a=0$, the zeroes and poles of the determinant are Casimirs, and 
the flow rescales the section $\phi$, so that choosing leaves for which 
the determinant is constant, and then rescaling so that  the determinant along 
the symplectic leaves is 1 lands us in $Sl(r,\bbc)$, giving us a space 
which is the reduction of the previous one up to a finite ambiguity:
$$\eqalign{(\mell)_{\rm red} =  \{ \phi  &{\rm \ meromorphic}\ 
Sl(r,\bbc)-{\rm valued\ functions\ on\ }\bbc\ {\rm\ such\ that}\cr
           & \phi^{-1}(\infty)\subset 
\cup_i\{\nu_i+\bbz\omega_1+\bbz \omega_2\} 
  {\rm\ and\ } \phi(\lambda+\omega_i) = I_i\phi (\lambda)I_i^{-1}, i=1,2\}. 
}\eqno(2.16)$$

{\it iii) Trigonometric case} 

This case can be thought of as a degeneration of the elliptic
case. The elliptic curve degenerates into a Riemann sphere $\Sigma= 
\bbp_1$ with two
points $z=0, \infty$ identified. The smooth part of the curve is then
$\bbc^*$. Let $\pi:\bbc\rightarrow \bbc^*$ be the map $z\mapsto exp(2\pi i 
z)$.

Let $D$ represent a sum $\nu_1+\nu_2+...+\nu_n$ of	
points $\nu_i$ on $\bbc^*$.

Our phase space $\mtrig$ will be the product of $\bbc^*$ with the space of 
$sl(r,\bbc)$-valued meromorphic
 functions $\phi$ on $\bbc$, with poles at $D$, satisfying
$$ \phi(\lambda+1) = I_1\phi(\lambda) I_1^{-1},\eqno(2.17)$$
$${\rm lim}_{{\rm Re}(-i\lambda)\rightarrow +\infty} \phi = 
{\rm lim}_{{\rm Re}(-i\lambda)\rightarrow -\infty} I_2\phi(\lambda) 
I_2^{-1}.\eqno(2.18)$$

As for the elliptic case, $\mtrig$ can be expressed as the space
$$\mtrig = \bbc^*\times  H^0(\Sigma, End(E)\otimes K_\Sigma(D)),\eqno(2.19)$$
for a suitable stable degree one vector bundle $E$ on $\Sigma$. The bundle is 
obtained
from the sum of line	bundles ${\cal O}\oplus	{\cal O}\oplus...\oplus{\cal 
O}\oplus{\cal O}(1)$
on $\bbp_1$ by identifying the fibers over $0,\infty$ in the standard 
trivialisations
by the matrix:
$$A= \pmatrix {0&0&0&\dots&0&1\cr 1&0&0&\dots&0&0\cr.&.&.& &.\cr.&.&.& 
&.\cr
\cr0&0&0&\dots&1&0}.\eqno(2.20)$$

As in the elliptic case, the extra $\bbc^*$ factor represents twists of $E$ 
by a line bundle; this translates here into a rescaling of the matrix $A$ 
by a scalar factor, and does not change the explicit expression of the 
endomorphisms. Exactly as in the elliptic case, one has 
$H^1(\Sigma, sl(E) )=  H^0(\Sigma, sl(E) )= 0$, and so projections 
$P_+,P_-$ and their 
difference $R$, as well as a projection $P_0$ onto the trace component.
Again, the line bundle ${\cal O}(D)$ has an $n$-dimensional space of 
sections, and
there is an $n+1$-dimensional family of	Poisson structures
on $\mtrig$ defined by the formula (2.13).
\bigskip

\noindent{\sc b) Spectral curves and line bundles.}

The three moduli spaces given above are particular examples of moduli 
spaces of Higgs pairs. 
For any compact Riemann surface $\Sigma$, and positive divisor $D$ of 
degree $n$ 
on $\Sigma$,
 one can consider [Ma], [Bo] the moduli space ${\cal H}(r,D,d)$ of {\it 
Higgs pairs} $(E, \phi)$, where
\item {-} $E$ is a degree $d$ rank $r$ holomorphic vector bundle over 
$\Sigma$.
\item{-} $\phi$, the {\it Higgs field},	is a holomorphic section of the 
associated adjoint bundle ${\rm End}(E)$,
twisted by $K_\Sigma(D)$, where $K_\Sigma$ is the canonical bundle of 
$\Sigma$:
$\phi\in H^0(\Sigma,{\rm End}(E)\otimes K_\Sigma(D))$. Alternately,
$\phi$ is a meromorphic ${\rm End}(E)$-valued 1-form, 
with poles at the divisor $D$. 

To each pair $(E,\phi)$, one can associate 
the {\it spectral curve} $S$ of $\phi$. This curve 
lies in the total space $T$ of the line bundle $K(D)$ over $\Sigma$.
It is cut out by the equation 
$$ \rm {det}(\phi(\lambda)-\zeta\bbi)=0.\eqno (2.21)$$
Here $\zeta$ represents the tautological section of $\pi^*K(D)$ over 
$T$,
where $\pi: T\rightarrow \Sigma$ is the projection. The projection
$\pi$ exhibits $S$ as an $r$-sheeted branched cover of $\Sigma$.
If $\gamma$ is the genus of $\Sigma$, the genus of $S$ is
$$g = r^2(\gamma-1) + {(r-1)rn\over 2} + 1.\eqno(2.22)$$

One can also define a sheaf $L$ supported on $\Sigma$, by the exact 
sequence over the surface
${\cal K}_D$: 
$$0\rightarrow \pi^*E\otimes K_\Sigma^*(-D)\quad 
{\buildrel{\phi-\zeta\bbi}\over
{\longrightarrow}}\quad \pi^* E\rightarrow L
\rightarrow 0 .\eqno(2.23)$$
On the generic locus for which $S$ is smooth and the eigenspaces are line 
bundles,
 $L$ will be a line bundle over $S$, of degree $d + r(1-\gamma) + g-1$.

One has:

{\sc Proposition }(2.24) [Hu] \tensl One can reconstruct $(E,\phi)$ from 
$(S,
 L)$:

\item {-} $E=\pi_*(L)$, 
\item {-} $\phi$ is, up to automorphisms,	the map induced on $E$ by 
multiplication by the tautological section $\zeta$
on $L$.\tenrm 

Our spaces ${\cal M}$ are all dense open sets in the moduli  ${\cal 
H}(r,D,d)$; the preceeding theorem  will allow us to describe the quotient 
${\cal N} = {\cal M}/Aut$  of 
${\cal 
H}(r,D,d)$ by a (constant over ${\cal M}$) group of automorphisms,
as a space of pairs $(S,L)$; alternately,
rather than consider pairs $(S,L)$, we note that when one thinks of 
$L$ as a sheaf over the surface ${\cal K}_D$, the curve $S$ is the support
of $L$, and so is specified by it. The space ${\cal N}$
 is then isomorphic to a space
of sheaves $L$ supported on  curves. The tangent space  
at $L$ is then given by the global Ext-group 
$$T{\cal N}_L = Ext^1(L,L),\eqno(2.25)$$
 and the cotangent space by 
$$T^*{\cal N}_L  = Ext^1(L,L\otimes K_T),\eqno(2.26)$$
whose computations are explained below. A Poisson structure is then given by a map
$$\Lambda:Ext^1(L,L\otimes K_T)\rightarrow Ext^1(L,L).\eqno(2.27)$$
It is the  fundamental observation of Mukai, Tyurin and Bottacin [Mu, Ty, 
Bo] that a Poisson structure
$\theta\in H^0(T, K_T^*)$ on the surface $T$ allows one to define such a 
structure,
 via the map $L\otimes K_T\rightarrow L$ that it induces.
We will see that the	surfaces which we consider have large families of 
Poisson structures,
and that	they correspond to the families of Poisson brackets given above.

 We consider each of our three cases.

{\it i) Rational case.} 

Here we take   bundles	of degree zero. The generic $(E,\phi)$ in this case
is such that $E$ is a trivial bundle; $\phi$ is then a matrix valued 
function of degree
$n$ with 
poles at $D$. One can multiply by a scalar polynomial, and take all poles
to be at infinity, normalising to $D= n\infty$. One then has 
$$\mrat\subset{\cal H}(r,D,0),\eqno(2.28)$$
 as  the open subset for which $E$ is trivial. On this subset, the 
automorphisms of $E$ are then given by 
constant matrices in $Gl(r,\bbc)$; under the isomorphism above, they act 
on $\mrat$ by conjugation, and if we let ${\cal N}_{\rm rat}$ be the variety of 
pairs $(S,L)$ corresponding to elements of ${\cal M}_{\rm rat}$, one has 
$$\mrat/Gl(r,\bbc) = {\cal N}_{\rm rat}.\eqno(2.29)$$
By (2.22), the genus of the spectral curves is 
$g = -r^2 + {(r-1)rn\over 2} + 1,$ and   the degree of the line bundles 
$L$ when the 
curve is smooth is $g+r-1$. The spectral curve is embedded in the total 
space $T$ of the line bundle
${\cal O}(n)$ over $\bbp^1$. The anticanonical bundle of $T$ is the lift 
from $\bbp^1$ 
of ${\cal O}(n+2)$. Poisson structures are then sections of this bundle: 
the space of Poisson structures
on $T$ is then of dimension $n+6$. If $\lambda$ is the standard coordinate 
on $\bbp^1$, and 
$z$ is a standard fiber coordinate corresponding to the tautological 
section $\zeta$ (so that $T$ is covered by two 
coordinate systems $(\lambda, z),
(\tilde\lambda,\tilde z)$ related by $(\tilde\lambda,\tilde z) = 
(\lambda^{-1},
 z\lambda^{-n})$), the	Poisson structures are given by 
$$ (a(\lambda) +b(\lambda) z ) ({\partial \over \partial\lambda} \wedge 
{\partial \over \partial z}),$$
where $a$ is a polynomial of degree at most $(n+2)$ and $b$ a polynomial 
of degree at most $2$.
We restrict to the $(n+3)$-dimensional space $W$ of polynomials $a, b$ of degrees at most $(n+1), 
0$, respectively; these Poisson structures on the 
surface  vanish over $\lambda = \infty$.

One has [AHH],[Sc],[HuMa]: 

 {\sc Proposition} (2.30) \tensl The Mukai structures that the Poisson 
structures in $W$ induce on the space $\mrat/Gl(n,\bbc)$
are the same as the reductions of the	Poisson structures given in section 
a).\tenrm

 As the references given are 
rather scattered and in some respects only partial, we give here a sketch 
of the proof, generalising
[HuMa]. 
The Mukai Poisson structure is given by a map from the cotangent space to 
the tangent space
$$\Lambda: Ext^1(L,L\otimes K_{\cal T})\rightarrow Ext^1(L,L).eqno (2.31)$$
To compute the $Ext$-groups, one must first take a resolution $R$ of $L$, 
take 
the induced sequence $Hom(R,L)$, and then compute the first 
hypercohomology group 
of this sequence, which we will do explicitly below. We choose the resolution of (2.23). Applying
$Hom$, and recalling that $K_{\cal T}= \pi^*{\cal O}(-n-2) $,
the cotangent space will be the first hypercohomology of the complex 
$C_{T^*}$ supported over the spectral curve
$$ (\pi^* E)^*\otimes L\otimes \pi^*{\cal O}(-n-2)\quad 
{\buildrel{(\phi-\zeta\bbi)^*}
\over
{\longrightarrow}}\quad (\pi^*E)^*\otimes L\otimes \pi^*{\cal O}(-2),\eqno (2.32)$$
and the tangent space the first hypercohomology of the complex $C_{T}$
$$ (\pi^* E)^*\otimes L \quad {\buildrel{(\phi-\zeta\bbi)^*}\over
{\longrightarrow}}\quad  (\pi^*E)^*\otimes L \otimes \pi^*{\cal 
O}(n).\eqno (2.33)$$
This is supported on the spectral curve. The map from the cotangent space 
to the tangent space
 is induced by termwise
multiplication in the resolution by the Poisson structure 
$(a(\lambda) +b(\lambda) z )({\partial \over \partial\lambda} \wedge 
{\partial \over \partial z}).$
Following [Ma], one can push down to $\bbp_1$, to have a diagram 
$$\matrix{ End(E)(-n-2)& {\buildrel{ad_\phi}\over
{\longrightarrow}} & End(E)(-2) \cr
\qquad \qquad\downarrow (a +b \phi)\cdot&&\qquad\qquad\downarrow (a +b 
\phi)\cdot\cr
End(E) & {\buildrel{ad_\phi}\over
{\longrightarrow}}& End(E)(n).}\eqno (2.34)$$
 The first hypercohomology of the top row is the cotangent space of 
${\cal N}_{\rm rat}$; that of the bottom row is the tangent space; the 
vertical maps induce the Poisson structure.

The first hypercohomology $\bbh^1$ of a sequence of sheaves $R{\buildrel{\rho}\over
{\longrightarrow}} S$ is given in terms of Cech cocycles (defined with 
respect to the standard cover
$U_+, U_-$ of $\bbp_1$) by equivalence classes of
pairs $(r_{\pm}, s_{+}, s_{-})$ where $r_{\pm}$ is a section of  $R$ over 
$U_+\cap U_-$,
$s_+, s_-$ are sections of $S$ over
$U_+, U_-$ respectively (a 0-cochain),
and $r,s$ satisfy $\rho(r_{\pm})-s_++s_-=0$ on $U_+\cap U_-$. The 
equivalence  
relation is that one can modify	$(r_{\pm}, s_{+},s_-)$ by  a coboundary $(\hat r_{+}-
\hat r_{-},
\rho(\hat r_{+}),\rho(\hat r_{-}))$ for 0-cochains $\hat r$ with values in 
$R$. 
In particular, one sees that the
hypercohomology group $\bbh^1$ maps to $ H^1(R)$ with 
$H^0(S)/\rho(H^0(R))$ in the kernel.

In our case, as the bundle is trivial, we have $H^1(\bbp_1, End(E)) = 0$: 
the bundle is rigid.
In particular, we have complementary projections $P_+,P_-$ from
$H^0(U_+\cap U_-, End(E))$ to $H^0(U_+, End(E))$, $H^0(U_-, End(E))$ 
respectively, which
coincide with the projections defined above.
Also, $	H^0(\bbp_1, End(E)) = 
gl (r,\bbc).$  If $C_T$ is the tangent complex in (2.34), this gives the isomorphism
$$H^0(\bbp_1, End(E)(n))/[\phi, gl(r,\bbc)]\rightarrow \bbh^1(C_{T}) 
=T{\cal N}_{\rm rat} ,\eqno(2.35)$$
Dually, we have $ H^0(\bbp_1, End(E)(-2))= 0,$
 $ H^1(\bbp_1, End(E)(-2))= gl (r,\bbc)^*$; furthermore,
again using Serre duality, the dual space to $H^0(\bbp_1, End(E)(n))$ is 
$H^1(\bbp_1, End(E)(-n-2))$. The phase space  $H^0(\bbp_1, End(E)(n))$  is 
identified with the space of matricial polynomials of degree at most $ n$, 
and dually, we represent elements of
 elements of $H^1(\bbp_1, End(E)(-n-2))$ by matrix valued Laurent 
polynomials with terms of degree $-n-1$ to $-1$; the Serre duality pairing 
is then trace-residue. Thus, if $C_{T^*}$ denotes the cotangent complex, 
there is an  isomorphism 
$$\bbh^1(C_{T^*}) =T^*{\cal N}_{\rm rat} \rightarrow(ker [\phi,\cdot]:	
H^1(\bbp_1, End(E)(-n-2))\rightarrow 
H^1(\bbp_1, End(E)(-2))).  \eqno(2.36)$$
Let us compute the Poisson tensor $\Lambda$.  An element of the cotangent 
space 
$T^*{\cal N}_{\rm rat}$ 
represented by a cocycle $c_\pm$ lifts to a hypercohomology cocycle
 $(c_{\pm}, d_+, d_-)$ with values in $End(E)(-n-2), End (E)(-2)$ 
respectively,
satisfying $[\phi,c_{\pm}] = d_+-d-$, so that one can take $d_+= 
P_+([\phi,c_{\pm}]), d_- =-P_-( [\phi,c_{\pm}])$.
 The Poisson tensor  $\Lambda$ acts by	
$$\Lambda: (c_{\pm}, d_{+}, d_-)\mapsto 
(a(\lambda)+b(\lambda)\phi)(c_{\pm}, d_{+},  d_-).\eqno(2.37)$$
We can modify the expression by a coboundary ${-b\over 
2}([\phi,c_{\pm}],[\phi,d_{+}],[\phi,d_{-}])$ so that the eventual 
expression for the Poisson tensor will be
more explicitly skew-symmetric:
$$\Lambda: (c_{\pm}, d_{+}, d_-)\mapsto ((a(\lambda)+{b(\lambda)\over 
2}\phi)(c_{\pm},  d_{+},  d_-)+
{b(\lambda)\over 2}(c_{\pm},  d_{+},  d_-)\phi).\eqno(2.38)$$

 As $H^1(\bbp_1, End(E)) = 0$,
we can split $a c_{\pm} + {b\over 2}(\phi c_{\pm} +  c_{\pm}\phi)$ using 
$P_+$, $P_-$, and modify 
our cocycle by the coboundary of $P_\pm(a c_{\pm} + {b\over 2}(\phi 
c_{\pm} +  c_{\pm}\phi))$. This gives 
the equivalent hypercohomology  cocycle 
$$\eqalign{ \big( 0, &\quad a P_+([\phi,c_\pm]) + {b\over 2}(\phi 
P_+([\phi,c_\pm])+ P_+([\phi,c_\pm])\phi ) 
- [\phi, P_+(a c_{\pm} + {b\over 2}(\phi c_{\pm} +  c_{\pm}\phi))],\cr
&-a P_-([\phi,c_\pm]) -{b\over 2}(\phi 
P_-([\phi,c_\pm])+P_-([\phi,c_\pm])\phi ) + 
[\phi, P_-(a c_{\pm} + {b\over 2}(\phi c_{\pm} +  c_{\pm}\phi))] \big) 
,}\eqno (2.39)$$
landing in the subspace $H^0(\bbp_1, End(E)(n))$, giving:
$$\eqalign{\Lambda: H^1&(\bbp_1, End(E)(-n-2))\rightarrow H^0(\bbp_1, 
End(E)(n))\cr
 c_\pm &\mapsto a P_+([\phi,c_\pm]) +{b\over 2}(\phi 
P_+([\phi,c_\pm])+P_+([\phi,c_\pm])\phi ) - [\phi, P_+(a c_{\pm} + {b\over 
2}(\phi c_{\pm} +  c_{\pm}\phi))]\cr 
=&-a P_-([\phi,c_\pm]) -{b\over 2}(\phi P_-([\phi,c_\pm])+ P_- 
([\phi,c_\pm])\phi ) + [\phi, P_-(a c_{\pm} + {b\over 2}(\phi c_{\pm} +  
c_{\pm}\phi))].}\eqno(2.40
)$$
Let us compute the Poisson bracket corresponding to this, on a pair of 
functions
$f, g$ on $ H^0(\bbp_1, End (E) (n))$. The differentials $df, dg$ of these 
functions at $\phi$
 are naturally identified
with classes in $H^1(\bbp_1, End(E)(-n-2))$. Recall that
 $Df=\phi\cdot df$, and $D'f=df\cdot \phi$, and similarly for $dg$.
We have 
$$\eqalign{\{f,g\}(\phi)&= <df, \Lambda(dg)>, \cr
                        &= <df,  a P_+([\phi,dg]) -  [\phi, P_+( a dg)]> 
\cr
                        &\quad + <{b\over 2} (Df+ D'f), P_+(Dg-D'g)>  
-<D'f-Df, P_+({b\over 2}
                        (Dg+D'g))>.}\eqno(2.41)$$
Using the identities 
$$\eqalign { <P_\pm(f), g> &= <f,P_\mp (g)>,\cr
<Df, Dg> &= <D'f, D'g>,}\eqno(2.42)$$ 
we obtain:
$$\eqalign { \{f,g\}(\phi) &= <\phi, [R(a\  df) , dg] +[df, R(a\ dg)]> 
\cr & \quad + {1\over 2}(<R(b(Df+D'f)), D'g-D g> + <Df-D'f, 
R(b(Dg+D'g))>,}\eqno(2.43)$$
which, when $b$ is a constant, reduces to  our brackets (2.5). 

{\sc Remark}: One might hope that the formula (2.5) would define a Poisson 
bracket for the structures not in $W$. Unfortunately, for cases like $a= 0, 
b= \lambda$, the Jacobi identity is not satisfied on the unreduced space. 
There does not seem to be any natural way of modifying the formula (2.5)
(i.e. its lift  to the unreduced space) so that the 
Jacobi identity is satisfied.

{\it ii) The elliptic case.}

The bundles $E'$ we consider over our elliptic curve are of degree 
one; on the  open set ${\cal M}_{\rm ell}$
of ${\cal H}(r,D,1)$ corresponding to the stable bundles, we have that 
$E' = E\otimes L'$, where $L'$ is a line bundle of degree zero and $E$ is 
the vector bundle defined above.
Any two such bundles $E', E''$ are isomorphic iff the corresponding line 
bundles $L',L''$
are such that  $(L'\otimes (L'')^*)^{\otimes r}$ is trivial, that is if 
$(L'\otimes (L'')^*$ is an $r$-th
root of unity in $Pic^0(\Sigma) \simeq \Sigma$ [At], so that the moduli 
space of bundles is then
the torus $\Sigma \simeq Pic^0(\Sigma)/(\bbz/r)^2$. The group $Aut$ of global 
automorphisms of the stable bundles are constant
multiples of the identity, 
so that ${\cal N }_{\rm ell} ={\cal M}_{\rm ell}/Aut = {\cal M}_{\rm ell} $.

By (2.22), the genus of the spectral curves is 
$g =  {(r-1)rn\over 2} + 1, $ and   the degree of the line bundles $L$ 
when the 
curve is smooth is	$g$. The spectral curve is embedded in the total space 
$T$ of the line bundle
${\cal O}(D)$ over $\Sigma$. The anticanonical bundle of $T$ is the lift 
from $\Sigma$ to $T$
of ${\cal O}(D)$. Poisson structures are then sections of this bundle: the 
space of Poisson structures
on $T$ is   of dimension $n+1$; the Poisson structures are of the form 
$$(a  + bz)({\partial \over \partial\lambda} \wedge {\partial \over 
\partial z}),\eqno(2.44)$$
where $a$ is a section of ${\cal O}(D)$ lifted from $\Sigma$, $b\in \bbc$ 
and $z$ is the tautological
 section
of $\pi^*({\cal O}(D))$ on $T$. We then have an $(n+1)$- dimensional family 
of Mukai brackets
 on ${\cal N }_{\rm ell}$.

{\sc Proposition (2.45)} [HuK, HuMa]\tensl
We have:
$$ \mell = {\cal N}_{\rm ell}.$$
The   Mukai Poisson structures on $ {\cal N 
}_{\rm ell}$ are
equivalent those on $\mell$ given above.\tenrm

{\sc Proof}: The identification of the spaces is given above; what remains 
to be done is to identify
the Poisson tensors. As for the rational case, we have a diagram:
$$\matrix{ End(E)(-D)& {\buildrel{ad_\phi}\over
{\longrightarrow}} & End(E)  \cr
\qquad \qquad\downarrow (a +b \phi)\cdot&&\qquad\qquad\downarrow (a +b 
\phi)\cdot\cr
End(E) & {\buildrel{ad_\phi}\over
{\longrightarrow}}& End(E)(D).}\eqno (2.46)$$
 The first hypercohomology of the bottom row is the tangent space
of  ${\cal N }_{\rm ell}$;  it decomposes as a sum
$$H^1(\Sigma, {\cal O}) \oplus H^0(\Sigma, End(E)(D)).$$  
The first factor corresponds to the tangent space of $\Sigma$ in $\mell$.
Dually, the cotangent space is a sum
$$H^1(\Sigma, End(E)(-D))\oplus H^0(\Sigma, {\cal O}).$$

We  recall  from above that since the bundles are rigid up to tensoring by 
a line bundle, we have 
$H^1(\Sigma, sl (E))= 0$, and so  
 we have complementary projections $P_+,P_-,P_0$ from
$H^0(U_+\cap U_-, End(E))$ to $H^0(U_+,sl(E))$, $H^0(U_-, sl(E))$, 
$H^0(U_+\cap U_-, {\cal O})$ respectively.

To compute the Poisson tensor, we take a cocycle
$c_\pm$ representing an element of $H^1(\Sigma, End(E)(-D))$, and a 
constant $c_E\in H^0(\Sigma, {\cal O})$. This corresponds to a 
hypercohomology 
cocycle
$$(c_\pm, d_+ + c_E\bbi, d_- + c_E\bbi)$$
where $d_+= P_+([\phi,c_{\pm}]), d_- =-P_-( [\phi,c_{\pm}])$.
 This gets mapped by $\Lambda$ to	
$(a +b \phi) (c_{\pm}, d_{+} + c_E\bbi, d_-+ c_E\bbi)$. 
We can split $(a +b \phi) c_{\pm}$ using $P_+$, $P_-$, $P_0$ and modify 
our cocycle by the coboundary of $\pm P_\pm((a +b  \phi) c_{\pm})$. This 
gives 
the equivalent cocycle 
$$\eqalign{(P_0((a +b \phi)c_\pm)\bbi,&\quad (a +b \phi)P_+([\phi,c_\pm]) -
 [\phi, P_+((a +b \phi) c_{\pm})] +(a +b \phi)c_E\bbi,\cr
&\quad  -(a+b \phi)P_-([\phi,c_\pm]) + [\phi, P_-((a+b\phi) c_{\pm})]+(a 
+b 
\phi)c_E\bbi),}$$
landing in the   $H^1(\Sigma, {\cal O})\oplus H^0(\Sigma, End(E)(D))$, 
giving:
$$\eqalign{\Lambda: H^1(\Sigma, End(E)(-D))\oplus H^0(\Sigma, {\cal O})
&\rightarrow H^1(\Sigma, {\cal O}) \oplus H^0(\Sigma, End(E)(D)),\cr
 (c_\pm, c_E)\qquad &\mapsto (P_0((a +b \phi)c\pm), \quad(a +b 
\phi)P_+([\phi,c_\pm])\cr
& -[\phi, P_+((a +b \phi) c_{\pm})]+(a +b \phi)c_E\bbi).}\eqno(2.47)$$
Let us compute the Poisson bracket corresponding to this, on a pair of 
functions
$f, g$ on ${\cal N }_{\rm ell}$. The differentials $df, dg$ of these functions at $\phi$
 are naturally identified
with classes   $(df_\phi, df_E), (dg_\phi, dg_E) \in H^1(\Sigma, 
End(E)(-D))\oplus H^0(\Sigma, 
{\cal O})$. Recall that
 $Df=\phi\cdot df$, and $D'f=df\cdot \phi$, and similarly for $dg$.
We have 
$$\eqalign{\{f,g\}(\phi)&= <df, \Lambda(dg)>\cr
                        &= <df_\phi,  a P_+([\phi,dg_\phi]) -  [\phi, P_+( 
a dg_\phi)] >,\cr
                        &\quad - b<   D'f_\phi, P_+(D'g_\phi)>  +b < 
Df_\phi, P_+( Dg_\phi)>\cr
                        &\quad  + < (df_\phi), (a+b\phi) dg_E\bbi>\cr
                        &\quad  - <df_E,  P_0((a+b\phi) dg_\phi)>.} 
\eqno(2.48)$$
Using the identities (2.42), again,
we obtain:
$$\eqalign { \{f,g\}(\phi) &= <\phi, [R(a\ df_\phi) , dg_\phi] +[df_\phi, 
R(a\ dg_\phi)]> 
\cr & \quad -{b \over 2}(<R(Df_\phi), Dg_\phi> + <D'f_\phi, R(D'g_\phi)>) 
\cr
&\quad + <P_0(a df_\phi +bDf_\phi), dg_E>- <df_E,  P_0((a  
dg_\phi+bDg_\phi)>,}\eqno(2.49)$$

which is the form of our brackets (2.13).

{\it iii) Trigonometric case}

This case is very similar to the elliptic case, and indeed is a limiting 
case of it.
On the nodal curve $\Sigma$ the stable bundles of degree one are 
classified by their top 
exterior power in $Pic^0(\Sigma)=\bbc^*$ [HuK]. The stable bundles, again, 
only have constant
 multiples of the identity as global automorphisms, so that again 
${\cal N }_{\rm trig} = {\cal M }_{\rm trig}/Aut = {\cal M }_{\rm trig}$, 
where ${\cal M }_{\rm trig}$ is the open subset of the moduli corresponding 
to stable bundles.
 
The only new element one must deal with comes from the singularity of the 
curve. One can work on the
desingularisation $\bbp_1$ of $\Sigma$. One-forms on $\Sigma$ are 
identified with one-forms on 
$\bbp^1$ with a simple pole at $0,\infty$ (and zero total residue), so 
that there is a 
global trivialisation of the cotangent bundle, given over $\bbp_1$ by the 
form $d\lambda/\lambda$;
dually, the tangent bundle is spanned by $\lambda{\partial\over 
\partial\lambda}$.

In a similar fashion, the Poisson structures on the total space $T$ of 
${\cal O}(D)$
correspond to Poisson structures on the   total space $\hat T$ of the lift 
of 
${\cal O}(D)$ to $\bbp_1$, of the form: 
$${\lambda\over p(\lambda)} (a(\lambda) + b z p(\lambda)) 
({\partial \over \partial\lambda} \wedge {\partial \over \partial z}),$$
where $p$ is a polynomial of degree $n$ vanishing at the lift of the 
divisor $D$, 
$a$ a polynomial of degree
$n$ satisfying $a(0)/p(0) = lim_{\lambda\rightarrow\infty} 
a(\lambda)/p(\lambda) $,
and $b$ is a constant. There is then an $n+1$-dimensional space of such 
structures on $T$.

As for the elliptic case, one can take the corresponding family of Mukai 
structures, and reduce them by the
action of $Pic^0(\Sigma)$, obtaining:

{\sc Proposition (2.50)} [HuK, HuMa]\tensl
We have:
$$ \mtrig= {\cal N}_{\rm trig} .$$
The   Mukai Poisson structures on
 ${\cal N}_{\rm trig}$ are
equivalent those on $\mtrig$ given above.\tenrm

\bigskip

\noindent{\sc c) Divisor coordinates and Poisson surfaces}

We have given in the preceeding section   (Poisson) embeddings  of 
spaces ${\cal M}/Aut$
into spaces ${\cal N}$ of pairs $(S,L)$, where $S$ is a spectral curve lying in a 
surface $T$
 and $L$ is a line bundle on the curve, (generically; in full generality,
$L$ is 
a sheaf on $T$  supported on $S$ ) . 

A line bundle on a curve can be represented by the divisor of zeroes of 
one of its holomorphic sections.
If the degree of the line bundle is equal to the genus of the curve, the 
line 
bundle  generically has a one-dimensional
space of sections, and the map which to a divisor of degree $g$ associates 
the corresponding
line bundle is a birational isomorphism between the $g$-th symmetric 
product $SP^g(\Sigma)$ 
of the curve and the variety $Pic^g(\Sigma)$.

The curve, however, lies in a surface, and so the divisor not only 
determines the line bundle, but also 
the curve, at least partially, as the curve must pass through  the points 
of the divisor. 

Over an open set of $T$ containing all the  spectral curves of an open set 
$U$ in our family, one can choose a 
line bundle $L_0$ such that the tensor product $L \otimes L_0$ is of 
degree $g$ over each 
spectral curve in $U$. Over an open set $V$  of pairs $(S, L)$, the line 
bundle $L\otimes L_0$
has a unique non-zero section, up to scale; this section vanishes over a 
divisor 
$\sum p_\mu $ of degree $g$. This divisor can be thought of as an element 
of the $g$th symmetric product
$SP^g(T)$ of $T$, or as a length $g$ zero-dimensional subscheme of $T$.

One straightforward way to obtain the divisor is to consider the defining 
sequence (2.23) for $L$:
$$0\rightarrow \pi^*E\otimes K_\Sigma^*(-D)\quad 
{\buildrel{\phi-\zeta\bbi}\over
{\longrightarrow}}\quad \pi^* E\rightarrow L
\rightarrow 0. $$
One can twist the fixed bundle $E$ so that it has up to scale a single 
section 
$\gamma$, which  then
by projection gives a section of 
$L$ which we will denote by $\gamma'$. The section   $\gamma'$ vanishes 
when   $\gamma$ lies in the image of 
$\phi-\zeta\bbi$. This gives the equation for the divisor
$$(\phi-\zeta\bbi)_{\rm adj}\gamma=0,\eqno (2.51)$$
where the subscript adj denotes the classical adjoint (matrix of 
cofactors).

As we have seen, the surfaces $T$ in which the spectral curves live have 
a family of Poisson structures, which are symplectic structures over open 
sets.
 These Poisson structures extend naturally to the symmetric product. The 
latter space is not smooth,
as it is singular over the diagonal. It does have a natural 
desingularisation, the Hilbert 
scheme $Hilb^g(T)$ of 0-dimensional ideals of length $g$, and the Poisson 
structures lift 
to $Hilb^g(T)$ [B]. 

{\sc Proposition (2.52)} \tensl For each Poisson structure in our family on 
$T$, taking the corresponding Mukai structure,
the map
$$\eqalign{ {\cal N}  &\rightarrow Hilb^g(T),\cr (S,L) &\mapsto \sum 
p_\mu}\eqno(2.53)$$
is Poisson. \tenrm

{\sc Proof:} The proof hinges on   the fact that the Mukai structures at 
$(S,L)$ are independent of the resolution
chosen of the sheaf $L$, and so one can choose a convenient resolution. 
Let us suppose we are at a generic point, at which $S$ is a smooth curve 
and 
$L$ a line bundle over $S$. If we extend $L$ to a neighbourhood of the 
curve, we have
$$ 0\rightarrow L\otimes N^*_S\quad 
{\buildrel{det(\phi-z\bbi)}\over{\longrightarrow }}\quad L \rightarrow 
L|_S\rightarrow 0$$
as a resolution. Taking $Hom(\cdot , L\otimes K_T)$ of this sequence, we 
have,
 that $T^*V $ is the 
first hypercohomology of the 
sequence
$$ {\cal O}_S\otimes K_T\rightarrow N_S\otimes K_T,$$
supported over $S$; the map is the zero map. $T^*V$ is then the sum
$$T^*V = H^1(S,{\cal O}_S\otimes K_T)\oplus H^0(S, K_S),\eqno(2.54)$$
 using the isomorphism $K_T\simeq K_S\otimes N^*_S$, and dually,
$$T V = H^0(S, N_S)  \oplus  H^1(S,{\cal O}_S ).\eqno(2.55)$$
 The pairing between $T^*V$ and $TV$ is given by Serre
duality:
$H^1(S,{\cal O}_S\otimes K_T)$ is the dual of $H^0(S, N_S)$ and 
$H^1(S,{\cal O}_S)$ is the dual of $H^0(S, K_S)$. The Poisson tensor 
applied to two covectors $(\alpha, \beta), (\alpha', \beta')$
is then given by 
$$<\alpha, \theta\beta'> - <\alpha', \theta\beta>.\eqno(2.56)$$
The proof is then a matter of writing out the pairings explicitly.
Let $\sum_\mu p_\mu$ be the divisor corresponding to $L$, and suppose for 
simplicity
that the points $p_\mu$ are distinct. Let $\rho= 0$ be the defining 
equation for the curve $S$,
and $\sigma$ be the section of $L\otimes L_0$, so that the $p_\mu$ are 
defined by the
 simultaneous vanishing of $\rho, \sigma$. On $TV$, variations of the 
curve $S$ 
 correspond to sections $v$ of the normal bundle $H^0(S, N_S)$, while 
variations of the line bundle 
are given by the cocycle $\dot \sigma/\sigma$ defined on punctured discs 
surrounding the $p_\mu$,
so that the differential $(\dot\sigma/\sigma, v) \mapsto \dot p_\mu$ 
of the map $(S,L)\rightarrow \sum_\mu p_\mu$ is given by the 
two conditions
$$\eqalign {d\rho(v(p_\mu))  &= d\rho( \dot p_\mu  ),\cr
\dot \sigma &= -d\sigma(\dot p_\mu ).}\eqno(2.57)$$

If $f,g $ are functions on $Hilb^g(T)$, let $F,G$ denote the corresponding 
functions
on ${\cal N} $. The differential $dF$ at a point $(S,L)$ is represented 
by a pair $(\tau^F, \omega^F)$
in the sum (2.54).
Representing $\tau^F$ by cocycles $\tau^F_\mu$ with values in $K_T$ on 
punctured disks around the $p_\mu$, we have
from (2.57) 
$$\tau^F_\mu  = d_\mu f\wedge {d\sigma\over \sigma},$$
while 
$$\omega^F(p_\mu)= d_\mu f|_S.$$ 
Evaluating the Poisson bracket on $F,G$.
$$\eqalign{ \{F,G\} &= < \theta \tau^F, \omega^G> - < \theta \tau^G, 
\omega^F>,\cr
                   &= \sum_\mu {\rm res}_\mu 
                          (<\theta, d_\mu f\wedge {d\sigma\over \sigma}> 
d_\mu g|_S -
                           <\theta, d_\mu g\wedge {d\sigma\over \sigma}> 
d_\mu f|_S ),\cr
                  &= \sum_\mu  
                          <\theta, d_\mu f\wedge d_\mu g >,}\eqno(2.58)$$
which is the Poisson bracket of $f,g$.

We note that for each Poisson structure in the family, 
 $Hilb^g(T)$ is symplectic over the open set of ideals whose 
support is disjoint from the zero divisor $P$ of the Poisson structure on 
$T$. The map which 
to a 
pair $(S,L)$ associates its divisor is generically immersive if one fixes 
$S$, or,
on the level of tangent spaces,   is generically injective on the summand 
$H^1(S,{\cal O})$ of $TV$. As the map is Poisson, this tells us that the
map ${\cal N} \rightarrow Hilb^g(T)$ is an isomorphism on the level of 
symplectic leaves.

{\it Reduction to $sl(r)$.}

In two of the cases which concerned us, the elliptic and trigonometric, 
the more usual 
phase spaces consist of $sl(r)$ or $Sl(r)$-valued functions, and, as we 
saw, we could obtain
 these phase spaces by reduction.
This was done by fixing the highest exterior
power of the bundle $E$, and then shifting $\phi$ either additively or 
multiplicatively by a constant so that it is  traceless or of 
fixed determinant. 
This has a good interpretation in the elliptic case, when $\Sigma$ is a 
group.
The top exterior power of $E$ is represented, up to a constant,
 by the sum $\sum_\mu \pi(p_\mu)$. 
Let us fix  a (linear) coordinate $\lambda$ on the base elliptic curve, 
and take a fiber
coordinate $z$ in $T$, such that the Poisson tensor on $T$ is given by 
$(a(\lambda) + bz) (
{\partial\over \partial \lambda}\wedge {\partial\over \partial z})$. 
Fixing the top exterior
power amounts to fixing $\sum_\mu \lambda_\mu$. Taking   $\sum_\mu 
\lambda_\mu$
as a Hamiltonian, we get flows $\dot \lambda_\mu = 0, \dot z_\mu =
 a(\lambda_\mu) + bz_\mu  $, which of course are compatible with the flows 
for $\phi$.
We can normalise using these flows; one generically valid normalisation is
$\sum_\mu z_\mu = 0$.

\bigskip

\noindent {\sc d) Integrable systems}

 There is a family of commuting  Hamiltonian systems defined on our different spaces, which 
indeed is integrable at levels (b), (c)  for all our Poisson structures. 
At level (a), it is given by Hamiltonians of the form
$$F_{\omega, n}(\phi) = res (\omega tr( \phi^n)),\eqno(2.59)$$
where $\omega$ is a scalar cocycle; in other words, the Hamiltonians are 
the coefficients of the spectrum of $\phi$.
At level (b), the Lagrangian leaves are given by fixing the spectral 
curve; in other words, they are the 
fibers of the projection
$$(S,L)\mapsto S.\eqno(2.60)$$
Corresponding to this, on the level of tangent spaces to ${\cal N} $, one has 
from
(2.55)  an exact sequence
$$0\rightarrow H^1(S,{\cal O})\rightarrow T{\cal N}\rightarrow  
H^0(S,N_S)\rightarrow 0.\eqno(2.61)$$
Indeed, deformations of line bundles on a spectral curve are given by 
$H^1(S,{\cal O})$, while deformations
of the spectral curve are given by sections of the normal bundle. It is 
quite easy to see that the 
foliation is Lagrangian, under the Mukai bracket. Indeed, 
  dually to (2.61), we have:
$$0\rightarrow H^1(S, K_T) \rightarrow T^*{\cal N}\rightarrow  
H^0(S,K_S)\rightarrow 0.\eqno(2.62)$$
Functions on the base of (2.60), lifted to ${\cal N}$, have their 
differentials in the summand
$H^1(S, K_T)$; under the action of the Poisson tensor $\Lambda$, this gets 
mapped to
$H^1(S,{\cal O})$. Referring to (2.56) for a pair of 
differentials $dF= (\alpha,0), dG= (\alpha', 0)$ 
of functions $F, G$ lifted from the base, we have
$$\{F,G\}=0,$$
so that the fibration is indeed Lagrangian. One can also see which of 
these functions are Casimirs:
they correspond to the kernel of the  map $H^1(S, K_T)\rightarrow 
H^1(S,{\cal O})$ 
given by multiplication by the Poisson tensor $\theta$ on $T$. This is 
part of a long exact sequence
$$ 0\rightarrow H^0(S, K_T)\rightarrow H^0(S,{\cal O}) \rightarrow  
H^0(P,{\cal O})
\rightarrow H^1(S, K_T)\rightarrow H^1(S,{\cal O})\rightarrow 0,\eqno 
(2.63)$$
where $P$ is the divisor of $\theta$ on the curve $S$. The image 
$\delta(H^0(P,{\cal O}))
\subset H^1(S, K_T)$ correspond to the differentials of the defining 
equations of the spectral
curve at $P$, and so these are the Casimirs. In other words, the 
symplectic leaves of the Poisson structure at level (b)
are given by fixing the spectral curves at $P$. We note that as $\theta$ 
varies, $P$ moves all over the 
spectral curve, and so:

{\sc Proposition} (2.67) \tensl Fixing the values of the Casimirs for all 
the Poisson structures 
in our families is tantamount to fixing the spectral curve $S$: the joint 
level sets of the
Casimirs is then  an open set of  the Jacobian of $S$. \tenrm

This proposition and the integrability of the system gives us Theorem 2 of 
the introduction. We note that at level (c) the lagrangian leaves in $Hilb^g(T)$ are given simply by 
constraining the points to lie
on the curves $S$, so that the leaves are $Hilb^g(S)\subset Hilb^g(T)$.

\bigskip

\noindent{\bf 3. Nijenhuis coordinates.}
\medskip

\noindent {\sc A) Privileged coordinate systems}
\medskip

Recapitulating, we have established a chain of  maps between:
\item {(a)} Certain spaces {\cal M} of matrix-valued functions on the 
line (times a curve, in the trigonometric and elliptic cases), equipped with a 
family of $r$-matrix-type
brackets;
\item{(b)} Spaces of pairs $(S,L)$ of spectral curves $S$ and sheaves $L$
 supported on these curves, equipped with a family of Mukai brackets;
\item {(c)} Spaces of length $g$ 0-dimensional ideals on a surface, 
equipped with a 
family of Poisson brackets, induced from a similar family on the surface.

The passage from (a) to (b) is obtained by thinking of the matrix valued 
function as a meromorphic 
endomorphism of a fixed vector bundle $E$, and taking then its spectral 
curve $S$
and its associated (dual) eigensheaf $L$. 

To summarise, we state the 

{\sc Theorem} (3.1) \tensl  On levels (a), (b), (c), we have linear 
families of Poisson structures
of dimensions $(n+3)$ in the rational case, $(n+1)$ in the elliptic and 
trigonometric 
cases. In each case, the 
families on the three levels can be identified, so that the maps relating 
levels 
(a), (b) and (c) are Poisson for each structure in the family. The map from 
level (a) to (b) takes a quotient by a group of automorphisms, which is 
trivial in the elliptic and trigonometric cases; the map from level (b) to 
(c) is generically immersive on symplectic leaves.  \tenrm

This is most of  Theorem 1 of the introduction. In case (c), because we are dealing with what is in essence a symmetric 
product of the surface, there 
are natural sets of coordinates which can be exploited. On a surface,  
Poisson structures 
are simply sections of the line bundle $K_T^*$, and so the quotient of 
any  two of them is a meromorphic function.
Thus, if we choose Darboux coordinates $\lambda, z$ for one of the 
structures in our family, so that the Poisson tensor is
${\partial \over \partial\lambda} \wedge {\partial \over \partial z}$, the 
other Poisson structures are
$f(\lambda, z){\partial \over \partial\lambda} \wedge {\partial \over 
\partial z}$, for $f$ a meromorphic function 
in a linear system. On the level of the symmetric product, then, the 
induced coordinates  $\lambda_\mu, z_\mu$
give a Poisson tensor of the form
$$ \Lambda = \sum_\mu f(\lambda_\mu, z_\mu){\partial \over 
\partial\lambda_\mu} \wedge {\partial \over \partial z_\mu}.\eqno (3.2)$$

Let us give the following  definition of Nijenhuis coordinates: suppose 
that one has two  Poisson tensors
$$\Lambda_1, \Lambda_2: T^*M\rightarrow TM,\eqno(3.3)$$
for some manifold $M$, and suppose that $ \Lambda_2$ is non-degenerate. We say that coordinates $f_1,...f_n$ on $M$ are 
Nijenhuis coordinates if their
differentials $df_i$ are eigenvectors of $\Lambda_2^{-1}\Lambda_1$. (This 
is  the  definition given, e.g. in [FP];  in the  next subsection we show 
how it is related   to that of [N].) One has:

{\sc Proposition} (3.4) \tensl The coordinates $\lambda_\mu, z_\mu$ are 
Nijenhuis coordinates in case 
(c) for any pair of Poisson structures in our family.\tenrm

One can ask what one does when the Poisson tensor has a kernel, for 
example at level (b). One 
frequent way of dealing 
with a map which has a kernel is to consider its graph instead, and 
generalise from there. Let us 
consider in 
$T^*M\times T^*M$ the space $V$: 
$$\{ (v_1, v_2) | \Lambda_1(v_1) = \Lambda_2(v_2)\},$$
and say that $v$ is an {\it eigenvector} of the pair $\Lambda_1, 
\Lambda_2$ if
$$ \Lambda_1(\alpha_1 v ) = \Lambda_2(\alpha_2 v ),$$
for some constants $\alpha_1, \alpha_2$, i.e., if $(\alpha_1 v,\alpha_2 
v )$ lies in $V$. With this convention, one can 
define Nijenhuis coordinates
in the degenerate case, as in the non-degenerate one. One has, for 
${\cal N} $ in the cases 
we have considered,
 
{\sc Proposition}(3.5) \tensl For any pair of Poisson structures 
$\Lambda_1, \Lambda_2$ 
in our family, if we  complete 
the functions  $\lambda_\mu, z_\mu$ by Casimir functions for either 
$\Lambda_1$ or 
$ \Lambda_2$ to a coordinate system, we obtain   Nijenhuis coordinates on 
${\cal N} $. \tenrm

\bigskip 

\noindent {\sc B) Definitions of  Nijenhuis coordinates.}

We now explain briefly how our definition of Nijenhuis coordinates, which we believe
is more adapted to a multi-Hamiltonian (as opposed to bi-Hamiltonian)
situation, is related to the classical definition of [N] ; see also  [Mg2], [GZ]. Given two compatible 
Poisson structures 
$$\Lambda_1, \Lambda_2: T^*M\rightarrow TM, \eqno (3.6)$$
with the second non-degenerate, the classical Nijenhuis coordinates [N] are defined as the eigenvalues of 
$\Lambda_1^{-1}\Lambda_2$. We shall see here that a more suitable notion 
when dealing with 
several Poisson structures is to consider  instead coordinate functions 
whose differentials are eigenvectors of $\Lambda_1^{-1}\Lambda_2$.

Let us suppose that the tensor $\Lambda_1^{-1}\Lambda_2$ can be 
diagonalised in a neighbourhood of a point with eigenvector 1-forms 
$\omega_i$ and eigenvalues $\rho _i$. A first remark is 

\noindent {\sc Lemma} (3.7) \tensl  Let $\rho_i\ne \rho_j$. Then 
$<\omega_i, \Lambda_1(\omega_j)> = <\omega_i, \Lambda_2(\omega_j)> = 0$ . 
\tenrm

\noindent {\sc Proof:} Recall that the Poisson tensors are skew adjoint. 
We then have
$$\eqalign { \rho_i <\omega_i, \Lambda_1(\omega_j)> &= 
<\Lambda_1^{-1}\Lambda_2(\omega_i), \Lambda_1(\omega_j)>,\cr
&= - <\omega_j,  \Lambda_2(\omega_i) >,\cr
&=  < \omega_i , \Lambda_2(\omega_j)>,\cr
&= - <\Lambda_1^{-1}\Lambda_2(\omega_j), \Lambda_1(\omega_i)>,\cr
&= - \rho_j<\omega_j, \Lambda_1(\omega_i)>,\cr
&= \rho_j<\omega_i, \Lambda_1(\omega_j)>.\cr}\eqno (3.8)$$
The proof for $\Lambda_2$ is similar.

The nondegeneracy of the forms then forces the eigenvectors to come in 
pairs, one pair for each eigenvalue. We will suppose, as a genericity 
constraint, that otherwise the eigenvectors are distinct. Rescaling the 
$\omega_i$ and renumbering the eigenvalues, we can then write the Poisson 
tensors as 
$$\eqalign{ \Lambda_1 = & \sum _{i=1}^n \omega^*_{2i-1}\wedge 
\omega^*_{2i},\cr
 \Lambda_2 = & \sum _{i=1}^n \rho_i\omega^*_{2i-1}\wedge 
\omega^*_{2i}.}\eqno (3.9)$$
and, dually, the symplectic forms as 
$$\eqalign{ \Omega_1 = & \sum _{i=1}^n \omega_{2i-1}\wedge \omega_{2i},\cr
 \Omega_2 = & \sum _{i=1}^n \rho_i^{-1}\omega_{2i-1}\wedge 
\omega_{2i}.}\eqno (3.10)$$
The next step is to show that one can replace the $\omega_{2i-1}\wedge 
\omega_{2i}$
by $dx_{2i-1}\wedge dx_{2i}$, for suitable coordinate functions. To do 
this, it will suffice to show that $d(\omega_{2i-1}\wedge \omega_{2i})=0$. 
Indeed, Frobenius' theorem then tells us that the 
distribution generated by $\omega_{2i-1}\wedge \omega_{2i}$ is integrable, 
and so one has 
functions $y_{2i-1}, y_{2i}$ such that $\omega_{2i-1}\wedge \omega_{2i} = 
f dy_{2i-1}\wedge dy_{2i}$, for some function $f$. But then, since 
$d(\omega_{2i-1}\wedge \omega_{2i})=0$, one finds that 
$f$ can only depend on $y_{2i-1}, y_{2i}$, and so one choose $x_{2i-1}= 
y_{2i-1}, x_{2i}= \int fdy_{2i} $.

\noindent {\sc Lemma} (3.11) $d(\omega_{2j-1}\wedge \omega_{2j})=0 , j = 
1,..,n$.

{\sc Proof:} We use the compatibility of the Poisson structures, more 
precisely, that 
 $\Lambda_1- a \Lambda_2$ is a Poisson structure for all values of $a$. 
This tells us that the forms 
$$ \Omega_a =  \sum _{i=1}^n (1-a\rho_i)^{-1}\omega_{2i-1}\wedge 
\omega_{2i},\eqno (3.12)$$
are closed for all values of $a$. Expanding in powers of $a$:
$(1-a\rho_i)^{-1} = 1 + a\rho_i + (a\rho_i)^2 +..$. and taking 
derivatives in $a$ at $a=0$ tells us that for all polynomials $p$, the 
forms 
$$ _p\Omega  =  \sum _{i=1}^n p(\rho_i) \omega_{2i-1}\wedge \omega_{2i}\eqno 
(3.13)$$
are also closed. Now, at a given point $x$, choose $p$ so that 
$p(\rho_j(x)) = 1, p(\rho_i(x))= 0$
for $i\ne j$, and $p'(\rho_i(x))= 0 $ for all $i$. At $x$, the exterior 
derivative of 
$ _p\Omega $ evaluates to $d(\omega_{2j-1}\wedge \omega_{2j})$ which must 
then vanish. As the choice of point is arbitrary, we are done. \hfill\za

Our symplectic forms are now:
$$\eqalign{ \Omega_1 = & \sum _{i=1}^n dx_{2i-1}\wedge dx_{2i},\cr
 \Omega_2 = & \sum _{i=1}^n \rho_i^{-1}dx_{2i-1}\wedge dx_{2i}.}\eqno 
 (3.14)$$
The last step is to remark that the fact that $\Omega_2$ is closed tells 
us that 
$\rho_i$ depends on $x_{2i-1}, x_{2i}$ only, giving
$$\eqalign{ \Omega_1 = & \sum _{i=1}^n dx_{2i-1}\wedge dx_{2i},\cr
 \Omega_2 = & \sum _{i=1}^n \rho_i^{-1}(x_{2i-1}, x_{2i}) 
dx_{2i-1}\wedge dx_{2i}.}\eqno (3.15)$$

We now have a normal form, and note that the  $x_i$ are Nijenhuis 
coordinates as their differentials are eigenvectors of the Nijenhuis 
operator. We have seen that under some mild non-degeneracy conditions, 
these coordinates exist, and come in pairs, one pair per eigenvalue. 
We note that the pair is defined only up to a volume preserving 
diffeomorphism of the plane. One canonical choice for the first coordinate 
of the pair, assuming some non-degeneracy,  would simply be the function 
$\rho_i(x_{2i-1}, x_{2i})$, giving the classical definition of 
Nijenhuis coordinate as an eigenvalue: the classical definition picks out 
one coordinate from amongst all of the functions of $x_{2i-1}, x_{2i}$.
On the other hand, with the modified definition, if we have an $n$-dimensional family of Poisson 
structures, one can show that with some non-degeneracy assumptions 
one has normal forms
$$\eqalign{ \Omega_1 = & \sum _{i=1}^n dx_{2i-1}\wedge dx_{2i},\cr
 \Omega_2 = & \sum _{i=1}^n \rho_{2,i}^{-1}(x_{2i-1}, x_{2i}) 
dx_{2i-1}\wedge dx_{2i},\cr 
\Omega_3 = & \sum _{i=1}^n \rho_{3,i}^{-1}(x_{2i-1}, x_{2i}) 
dx_{2i-1}\wedge dx_{2i},\cr&.\ .\ .\cr 
\Omega_n = & \sum _{i=1}^n \rho_{n,i}^{-1}(x_{2i-1}, x_{2i}) 
dx_{2i-1}\wedge dx_{2i},\cr
}\eqno (3.16)$$
so that the coordinates $x_i$ are Nijenhuis coordinates for the whole 
family, i.e., independently of the pair $\Omega_i, \Omega_j$ one chooses.

\noindent {\bf 4. More explicit formulae: the rational case.}

We now explain how the above theorems give coordinate systems which are 
quite tractable computationally, and so allow fairly explicit integration 
of the systems. We will simplify, and only consider the  rational case. 
The necessary extra computations required to deal with the elliptic and 
trigonometric cases are given in [HuK].

Recall that our level (a)  phase space, consisting of matricial polynomials 
$\phi(\lambda)$ of degree at most $n$, reduces at level (b) to a space of 
generically smooth spectral curves defined by 
$$det(\phi-z\bbi) = 0, $$ 
and (generically)  line bundles $L$ defined by 

 $$0\rightarrow {\cal O}^{\oplus r} \otimes {\cal O}(-n)\quad 
{\buildrel{\phi-z\bbi}\over
{\longrightarrow}}\quad {\cal O}^{\oplus r}\rightarrow L
\rightarrow 0 .\eqno(4.1)$$

The divisor coordinates at level (c) are obtained by considering the 
points on $S$ where the projection to $L$ of a standard section of 
${\cal O}^{\oplus r}$ vanishes. We choose the section $\gamma = (1,0,...,0)^T$
of ${\cal O}^{\oplus r}$; as a section of $L$, it vanishes where it lies 
in the image of ${\phi-z\bbi}$, that is, when
$$ (\phi-z\bbi)_{\rm adj} \gamma = 0,\eqno (4.2)$$
where the subscript denotes taking the matrix of cofactors (classical adjoint).
If we assume that $\phi$ is normalised to have its leading order term 
diagonal, this gives $r-1$ fixed (for a given spectral curve) points 
over $z = \infty$, and $g={\rm 
genus}(S)$ points $p_\mu = (\lambda_\mu, z_\mu)$ 
over the rest of the curve. It is these points that provide the 
coordinates. Another way of obtaining these points, following [Sc, Ge] is by noting that 
$\lambda_\mu$ are the $\lambda$-coordinates of points for which $V$ is not a cyclic 
vector for $\phi(\lambda)$, and so a defining equation for $\lambda_\mu$ is 
$$ det (V, \phi(\lambda)(V), \phi(\lambda)^2(V),..., 
\phi(\lambda)^{r-1}(V)) = 0.\eqno (4.3)$$
One can then obtain the $z_\mu$ as follows: one chooses another vector $W$ 
and sets
$$ \eqalign{P(\lambda) = & det (W, V, \phi(\lambda)(V), \phi(\lambda)^2(V),..., 
\phi(\lambda)^{r-3}(V),\phi(\lambda)^{r-2}(V)),\cr
R(\lambda) = & det (W, V, \phi(\lambda)(V), \phi(\lambda)^2(V),..., 
\phi(\lambda)^{r-3}(V),\phi(\lambda)^{r-1}(V)),\cr}\eqno (4.4)$$
One chooses $W$ so that $P(\lambda_\mu)$ is non-vanishing. One then has:
$$z_\mu = (-1)^r  [{R(\lambda_\mu)\over P(\lambda_\mu)} - {\rm 
tr}(\phi(\lambda_\mu))].\eqno (4.5)$$

This  then gives us the Nijenhuis coordinates in a fairly explicit way. 
We now consider the Poisson structures. As noted above, at level (c), we 
have a family of Poisson structures of the form
$$\Lambda_{a,b} = \sum _\mu (a(\lambda_\mu) +b z_\mu ) ({\partial \over \partial\lambda_\mu} \wedge 
{\partial \over \partial z_\mu}),\eqno(4.6)$$
where $a$ is a polynomial of degree at most $(n+1)$ and $b$ a constant.
 Choosing a polynomial $\alpha$ of degree at most $n$, one can consider the 
 two-dimensional linear subfamily 
(pencil) of Poisson structures
$$\Lambda_{c, c'} = \sum _\mu (c\lambda_\mu-c') \alpha(\lambda_\mu) 
({\partial \over \partial\lambda_\mu} \wedge 
{\partial \over \partial z_\mu}),\eqno(4.7)$$
where $c, c'$ are constants. Notice then that the $\lambda_\mu$ are  (classical) Nijenhuis coordinates 
for the Poisson structures $\Lambda_{1,0}, \Lambda_{0,1}$. The Casimirs for 
the $\Lambda_{1,c}$ structure include the intersections of the spectral 
curve with $\lambda = c$, and so are the $z$-coefficients of 
$$det(\phi(c)-z\bbi)=0.$$
These are polynomials of degree at most $nr$ in $c$, giving our 
generalisation of the Gel'fand-Zakharevich theorem. We note that as 
$c$ varies, the whole spectral curve is swept out, and so one indeed 
has the full set of Hamiltonians from taking the union of the Casimirs over 
all $c$.

As shown in [AHH], the flows corresponding to the Poisson structure  
  $\Lambda_{a,b}$ and the Hamiltonian $H_i$ (choosing a basis $H_1,..,H_k$
for the space of Hamiltonians)  can be obtained as follows, through a 
fairly standard generating function argument. Fixing the 
Hamiltonians fixes the spectral curve, and so  determines $z$ as a function of 
$\lambda$: $z = z
(\lambda,H_1,...,H_k)$.
Choosing a base point $\lambda_0$ on the spectral curve, we set  
$$\eqalign {F(\lambda_1,...,\lambda_g, H_1,...,H_k) =& \sum _\mu \int_{\lambda_0}
^{\lambda_\mu} b^{-1}{\rm ln}( a(\lambda) - bz(\lambda, 
H_1,...,H_k))d\lambda, \ {\rm for }\ b\ne 0,\cr
=& \sum _\mu \int_{\lambda_0}
^{\lambda_\mu}   ( a(\lambda))^{-1}z(\lambda, 
H_1,...,H_k))d\lambda, \ {\rm for }\ b= 0.}
\eqno (4.8)$$
The linearising
coordinates 
of the flows are 
given by 
$$Q_i = {\partial F\over \partial H_i} = \sum _\mu \int_{\lambda_0}^{\lambda_\mu}
( a(\lambda) - bz)^{-1}
{\partial \lambda\over \partial H_i}dz. \eqno (4.9)$$
One can show that these are sums of Abelian integrals.

One then has explicit formulae for the flows. In solving a particular 
system, it is then   a matter of writing out the separating variables 
in terms of the variables at hand. We give briefly the example of 
 the Neumann oscillator, 
describing motion on the sphere $\sum _{i=1}^nx_i^2 = 1$ under the 
influence of a quadratic potential $\sum _{i=1}^n\alpha_ix_i^2$, with 
conjugate momentum variables $y_i$ satisfying  $\sum _{i=1}^nx_iy_i = 0$.
(Compare [P]).
One has, at the loop algebra level, the following parametrization:

We set 
$$a(\lambda)  =
\prod_{i=1}^n(\lambda-\alpha_i), \quad a_j(\lambda)  =
\prod_{i=1, i\ne j}^n(\lambda-\alpha_i),\eqno(4.10)$$
and set 

$$\phi(\lambda) = a(\lambda) \pmatrix {0, & -1/2\cr
                             0, & 0 }
+  
\pmatrix{
-\sum_{i=1}^n  {x_iy_i}a_i(\lambda),
&-\sum_{i=1}^n {y_i^2}a_i(\lambda)\cr \cr
 \sum_{i=1}^n {x_i^2}a_i(\lambda), 
&  \sum_{i=1}^n {x_i y_i}a_i(\lambda)}.
\eqno (4.11)$$
The Hamiltonian for the system is 
$$Res_\infty ( det (\lambda 
\phi(\lambda) / a(\lambda)^2.\eqno (4.12)$$
The separating coordinates are simply the roots $\lambda_\mu$ of 
$$ \sum_{i=1}^n {x_i^2}a_i(\lambda) = 0,\eqno(4.13)$$
(these are the classical ellipsoidal coordinates) together with 
$$z_\mu = \sum_{i=1}^n  {x_iy_i}a_i(\lambda_\mu).\eqno(4.14)$$

\noindent {\bf References}
  
\item{[At]}
M. Atiyah,
{\it Vector bundles over an elliptic curve},
Proc. Lond. Math. Soc {\bf 7}, 414--452 (1957).

\item{[AHH]}
M.R. Adams,J. Harnad, and J. Hurtubise, 
{\it Darboux coordinates and Liouville-Arnold integration in loop 
algebras}, 
Comm. Math. Phys. {\bf 155} , no.~2, 385--413 (1993).

\item{[AHP]} 
M.R. Adams, J. Harnad, and E. Previato, 
{\it Isospectral Hamiltonian flows in finite and infinite dimensions {\rm 
I}.
Generalised Moser systems and moment maps into loop algebras}, 
Comm. Math. Phys. {\bf 117}, no.~3, 451--500 (1988).

\item{[AvM]}
M. Adler and P. van Moerbeke, 
{\it Completely integrable systems, Euclidean Lie algebras, and curves}, 
Adv. in Math. {\bf 38}, no.~3, 267-317 (1980); 
{\it Linearization of Hamiltonian systems, Jacobi varieties and
representation theory}, ibid. {\bf 38}, no.~3, 318--379 (1980).

\item {[B]} 
 A.Beauville,   {\it Vari\'et\'es K\"ahl\'eriennes 
dont la premi\`ere classe de Chern est nulle}. Jour. Diff.Geom. {\bf 18} 
755-782  (1983).

\item{[Bo]}
F. Bottacin, 
{\it Symplectic geometry on moduli spaces of stable pairs,}
Ann. Sci. Ecole Norm. Sup. (4) {\bf 28}, no. 4, 391-433 (1995).

\item {[FP]} G. Falqui and M. Pedroni, {\it Separation of variables for 
bi-Hamiltonian systems},  47 pages, preprint.

\item{[FT]}
L.D. Faddeev and L.A. Takhtajan,
{\it Hamiltonian methods in the theory of solitons},
eds., Springer-Verlag, Berlin, 1987.

\item {[Ge]} M.I. Gekhtman,
{\it Separation of variables in the classical ${\rm SL}(N)$ magnetic 
chain.}  
Comm. Math. Phys. {\bf 167}, no. 3, 593--605 (1995).

 \item{[GZ]} I.M. Gel'fand and I. Zakharevich, {\it On the local geometry of
a bi-Hamiltonian structure} in: The Gel'fand Mathematical Seminars 1990-1992
(L. Corwin et al. eds.) Birkhauser, Boston 1993, pp. 51-112.

\item{[HHu]} 
J. Harnad and J. Hurtubise, 
{\it Generalised tops and moment maps into loop algebras}, 
J. Math. Phys. {\bf 37}, no.~7, 1780--1787 (1991).

\item{[Hi1]} 
N.J. Hitchin, 
{\it The self-duality equations on a Riemann surface}, 
Proc. London Math. Soc. (3) {\bf 55}, no.~1, 59--126 (1987). 

\item{[Hi2]} 
N.J. Hitchin,  
{\it Stable bundles and integrable systems}, 
Duke Math. J. {\bf 54}, no.~1, 91--114 (1987).

\item {[Hu]}
J. Hurtubise,
{\it Integrable systems and algebraic surfaces},  
Duke Math. J. {\bf 83}, no.~1, 19--50 (1996).

\item {[HuK]} J. Hurtubise and M. Kjiri, {\it Separating coordinates for 
the generalized Hitchin systems
and the classical r-matrices}  Commun. Math. Phys {\bf 210}, 521-540 (2000).

\item{[HuMa]} J. Hurtubise and E. Markman, {\it Surfaces and the Sklyanin 
bracket}, 19 p., to appear in  Commun. Math. Phys. math.AG/0107010.

\item{[Ma]}
E. Markman,   
{\it Spectral curves and integrable systems}, 
Compositio Math. {\bf 93}, 255-290, (1994).

\item{[Mg1]} F. Magri, ``Eight lectures on Integrable Systems''  {\it  
Integrability of nonlinear systems }(Pondicherry, 1996), 
Lecture Notes in Phys., {\bf 495} Springer, Berlin-Heidelberg 1997, 256--296.

\item{[Mg2]} F. Magri, ``Geometry and Soliton Equations'' {\it La m\'ecanique 
analytique de Lagrange et son h\'eritage}, Atti Acc. Sci. Torino Suppl. 
{\bf 124}, 181-209 (1990).

\item{[Mo]} J.  Moser,  ``Geometry of Quadrics and Spectral Theory", 
{\it The Chern Symposium, Berkeley, June 1979}, 147-188, Springer, New 
York, 1980.

\item{[Mu]} S. Mukai,  {\it Symplectic structure of the moduli space of 
sheaves on an 
abelian or $K3$ surface.} Invent. Math. {\bf 77}, no. 1, 101--116 (1984).

\item{[N]} A. Nijenhuis, {\it $X_{n-1}$-forming sets of
eigenvectors.} Nederl. Akad. Wetensch.
Proc. Ser. A. {\bf 54} -- Indagationes Math. {\bf 13},
200--212 (1951).

\item {[NS]} M. Narasimhan and C.S. Seshadri, 
{\it Stable and Unitary vector bundles on a compact Riemann surface}, 
Annals of Math {\bf 82}, 540-567 (1965).

\item {[P]} M. Pedroni, {\it Bi-Hamiltonian aspects of the separability 
of the Neumann system} 
12 pages, nlin.SI/0202023 .

\item{[RS1]} 
A.G. Reiman and M.A. Semenov-Tian-Shansky, 
{\it Reduction of Hamiltonian systems, affine Lie algebras and lax 
equations {\rm I, II}}, 
Invent. Math. {\bf 54}, no.~1,  81--100 (1979); ibid. {\bf 63}, 
no.~3, 423--432 (1981).

\item{[RS2]} 
A.G. Reiman and M.A. Semenov-Tian-Shansky, 
{\it Integrable Systems II}, chap.2, in ``Dynamical Systems VII'', 
Encyclopaedia of Mathematical Sciences, vol 16., 
V.I. Arnold and S.P.Novikov, eds., Springer-Verlag, Berlin, 1994.

\item{[Sc]}
D.R.D. Scott, 
{\it Classical functional Bethe ansatz for $SL(N)$:separation of
variables for the magnetic chain}, 
J. Math. Phys. {\bf 35}, 5831-5843 (1994).

\item{[Sk1]}
E.K. Sklyanin,
{\it On the complete integrability  of the Landau-Lifschitz equation},
LOMI preprint E-3-79, (1979).

\item{[Sk2]}
E.K. Sklyanin,
{\it Poisson structure of a periodic classical $XYZ$-chain},
J. Sov. Math. ,{\bf  46}, 1664-1683  (1989) .

\item{[Ty]} A.N. Tyurin, 
{\it Symplectic structures on the moduli spaces of vector bundles 
on algebraic surfaces with $p\sb g>0$. } (Russian) 
Izv. Akad. Nauk SSSR Ser. Mat. {\bf 52}, no. 4, 813--852, 896  (1988);
 translation in Math. USSR-Izv. {\bf 33}, no. 1, 139--177  (1989). 

\bigskip

\medskip
\noindent J.Harnad: {\it Department of Mathematics and Statistics, 
Concordia University, and Centre de Recherches Math\'ematiques.
(harnad@crm.umontreal.ca)}

\noindent J. C. Hurtubise: {\it  Department of Mathematics and Statistics, 
McGill University, and Centre de Recherches Math\'ematiques. 
(hurtubis@crm.umontreal.ca)

\end

$$ \eqalign{ H^1(S,{\cal O}_S )\oplus H^0(S, N_S) &\rightarrow \oplus_\mu 
T_{p_\mu}T\cr
\dot \sigma/\sigma, \dot \rho &\mapsto -\pmatrix{d\sigma\cr d\rho}^{-1} 
\cdot \pmatrix
{\dot \sigma \cr \dot \rho}( p_\mu)}\eqno (??)$$

Another possibility, considering all the Poisson structures at once, is to 
take all the Hamiltonians $H_i$
of the integrable system as coordinates: these are the components of the 
defining equation 
 $det (\phi-z\bbi)=0$ of the spectral curve; to these one can add just 
the coordinates $\lambda_\mu$, 
which are Nijenhuis coordinates for all pairs of Poisson structures in the 
system, to obtain a full set of coordinates.

The map quotients out the 
automorphisms of $E$.
The passage from (b) to (c) is then given by by taking the divisor 
representing $L$; the spaces in (c) are
generically symplectic, and the map from (b) to (c) is a birational 
isomorphism from
 the symplectic leaves of (b) to the space of (c), for each of the Poisson 
structures.

equations of motion for the Hamiltonian system are equivalent to
$$
{d\phi\over  dt} = [ B, \phi],
\eqno(4.12)
$$
where
$$
  B = \pmatrix{
\sum_{i=1}^n  {x_iy_i}, &\lambda +\sum_{i=1}^n {y_i^2}\cr \cr
-\sum_{i=1}^n {x_i^2}, &-\sum_{i=1}^n  {x_iy_i}
 }.\eqno(4.13)
$$